\documentclass[twocolumn,footinbib,prb,superscriptaddress]{revtex4-2}
\usepackage{color}
\usepackage{amsfonts,amssymb,amsmath,mathrsfs}
\usepackage{amsbsy}
\usepackage{graphicx}
\usepackage{hyperref}

\begin{document}

\title{Robust quantum gates using smooth pulses and physics-informed neural networks}

\author{Utkan G\"ung\"ord\"u}
\altaffiliation[Current address: ]{Laboratory for Physical Sciences, College Park, Maryland 20740, USA}
\email{utkan@lps.umd.edu}
\affiliation{Department of Physics, University of Maryland Baltimore County, Baltimore, MD 21250, USA}

\author{J.~P.~Kestner}
\affiliation{Department of Physics, University of Maryland Baltimore County, Baltimore, MD 21250, USA}

\begin{abstract}
The presence of decoherence in quantum computers necessitates the suppression of noise. Dynamically corrected gates via specially designed control pulses offer a path forward, but hardware-specific experimental constraints can cause complications. Existing methods to obtain smooth pulses are either restricted to two-level systems, require an optimization over noise realizations or limited to piecewise-continuous pulse sequences. In this work, we present the first general method for obtaining truly smooth pulses that minimizes sensitivity to noise, eliminating the need for sampling over noise realizations and making assumptions regarding the underlying statistics of the experimental noise. We parametrize the Hamiltonian using a neural network, which allows the use of a large number of optimization parameters to adequately explore the functional control space. We demonstrate the capability of our approach by finding smooth shapes which suppress the effects of noise within the logical subspace as well as leakage out of that subspace.
\end{abstract}

\maketitle

\section{Introduction}
There has been significant progress in the past decades towards the realization of a physical quantum computer. The greatest obstacle presently hampering the current efforts, particularly for solid-state qubits such as Josephson junction based qubits or semiconductor spin qubits, is the presence of unwanted couplings between the qubit and its hosting environment \cite{Paladino2014}. An established way of suppressing the effects of such non-Markovian noise (as well as calibration errors in the Hamiltonian) since the early days of NMR is dynamical correction \cite{Vandersypen2005} ---the application of carefully designed control fields such that the effect of low-frequency noise can be arranged to cancel out when integrated over the entire evolution. This approach is especially relevant for noisy intermediate-scale quantum (NISQ) devices \cite{Preskill2018} whose limited number of qubits precludes implementation of quantum error correction codes, which can require as many as thousands of physical qubits to protect just one logical qubit \cite{Fowler2012}. Dynamical correction instead addresses noise without any overhead in the number of qubits at the cost of lengthening and complicating the control pulses.
For a robust quantum control protocol to be practical, however, it is important to take the limitations of the control hardware into account. Experimentally realistic control fields must have bounded amplitude and bandwidth, and hence be smooth pulses \cite{Barnes2015,Machnes2018}.

For an ideal qubit with no noise, smooth pulses can be obtained analytically \cite{Barnes2012} or numerically \cite{Machnes2018}. For dynamical correction, smooth pulses that are robust against errors within a qubit's logical subspace have also been proposed to address quasistatic noise in two-level systems \cite{Barnes2015,Zeng2019,Gungordu2019b} by reducing sensitivity to noise in a perturbative analysis.
Extending these (semi-)analytic approaches beyond two-level systems remains a challenge.
This is an important problem because although two-level systems are useful for basic demonstrations, a quantum computer requires at least pairwise interaction of qubits, entailing dynamics in a larger Hilbert space. Furthermore, even for single qubit control, systems such as the transmon \cite{Motzoi2009} and resonator-coupled spin qubits \cite{Warren2019} contain not only pairs of low lying energy levels used to encode a qubit, but also slightly higher lying ``leakage" states, requiring navigation of a larger Hilbert space that is outside the scope of these robust smooth pulses.

Other existing approaches numerically search for pulses that maximize gate fidelity for an ensemble of simulated noise sampled from an assumed distribution function, and are not restricted to two-level systems. The control is typically (although not necessarily \cite{Nobauer2015}) split into piecewise-constant segments \cite{Dong2016,Zahedinejad2016,Palittapongarnpim2017,Bukov2018,Niu2019,Ge2020,Wu2019}. In the piecewise-continuous approach, when the number of segments is not sufficiently large, a secondary optimization may be required to account for the filtering effects of the bandwidth-limited waveform generator on the sudden jumps in the pulse \cite{Machnes2018}. Although one may strive to approach a smooth pulse by increasing the number of segments and penalizing sudden jumps at the cost of computational complexity, in our work we instead use continuous functions which enable the efficient use of adaptive-time ordinary differential equation (ODE) solvers.
For broadband noise, a smooth pulse approach to improve gate fidelity by suppressing the leading order effects of an assumed noise power spectral density (PSD) was reported in Ref.~\onlinecite{Huang2017}. A common theme in all these approaches is that the optimization of fidelity is performed over the combined effect of the noise and control, which requires assumptions regarding the spectrum or the statistics of the noise and averaging over noise realizations.

Here, we propose a new approach that addresses all these issues simultaneously and allows us to find \emph{smooth} pulses for implementing robust quantum gates in any qubit platform. The generic approach described here is not limited to two-level systems and can suppress the effects of both leakage and quasistatic noise, regardless of the underlying statistics of the noise, by minimizing the \emph{sensitivity} to arbitrary quasistatic noise (as opposed to maximizing fidelity for a given noise sampling) through the exact functional relation between noise and its effect on the resulting quantum gate. This is enabled by recently introduced physics-informed neural network (PINN) frameworks \cite{Raissi2019} which implement deep neural networks (DNNs) that can be used to minimize a cost function while at the same time respecting a given set of differential equations (see also Refs.~\onlinecite{Psichogios1992,Lagaris1997,Lagaris1998}). Furthermore, as we will show, the resulting robust pulses are practical and constitute a new state-of-the-art in quantum control.

From a computational point of view, sampling-based approaches which focus on maximizing the gate fidelity for a given stochastic perturbation require additional optimization cycles with different noise realizations whose values are sampled from a given distribution function, which comes at a significant computational cost. By focusing on the noise sensitivity, our approach eliminates the necessity for averaging over noise traces. Furthermore, operating over smooth functions rather than a large number of piecewise-continuous pulses with fixed-width segments allows additional speed up when using adaptive ODE solvers \cite{Machnes2018}.
Another advantage of our functional approach is that one can impose exact boundary conditions and functional constraints (such as explicit bandwidth constraints, maximum drive amplitude, boundary conditions for the drive, or even functional constraints such as DRAG \cite{Motzoi2009} and filtering effects \cite{Motzoi2011}) from the outset. This is in contrast with the usual approach of penalizing any deviation of such constraints by including them in the cost function, which makes it more difficult to find solutions by making the optimization landscape more complicated, leading to additional false minima and stiffness. Furthermore, as we demonstrate below, it is also feasible to suppress higher order effects of noise within our approach, which are important for strong errors and long pulse durations.

The parameterization of the smooth control Hamiltonian using a DNN, instead of the commonly used Fourier harmonics amplitudes \cite{Nobauer2015,Huang2017,Machnes2018}, is a distinguishing feature of our approach which enables the use of a very large number of optimization parameters to exhaustively probe the functional space of smooth control fields. This is impractical with a Fourier series, since increasing the number of harmonics necessarily introduces faster oscillations which slow down adaptive ODE solvers, and no method to compute the parameter gradients with improved efficiency  (such as the backpropagation algorithm for DNNs) exists.

Finally, in terms of methodology, within the wider context of DNN based quantum control \cite{Leung2017,Fosel2018,Abdelhafez2019,Ban2020}, our approach is a new direction beyond sampling based learning protocols for finding robust and nonrobust shaped pulses.

\section{Method}
We start from a generic definition of a quantum system described by a control and error Hamiltonian as $H(t, \boldsymbol p) = H_c(t, \boldsymbol p) + H_\epsilon(t, \boldsymbol p)$ with
\begin{align}
H_c(t, \boldsymbol p) = \sum_i h_i(t, \boldsymbol p) \Lambda_i, \quad H_\epsilon(t, \boldsymbol p) = \sum_{i}  \epsilon_i \chi_i(t, \boldsymbol p) \Lambda_i,
\end{align}
where $\boldsymbol p = \boldsymbol p(t)$ is a time-dependent control parameter characterizing the driving fields $h_i$, $\epsilon_i$ are the quasistatic stochastic noise strengths, $\Lambda_i$ are traceless generators of the Lie algebra $\mathfrak{su}(n)$ that obey $\text{tr}(\Lambda_i \Lambda_j) = n \delta_{ij}$, and $\chi_i(t,\boldsymbol p)$ represent any dependence of the noise Hamiltonian $H_\epsilon$ on the control fields.  The latter becomes relevant when a term in the Hamiltonian is a nontrivial function of a noisy parameter, e.g., single-qubit microwave driving $\Omega(t)$ with multiplicative amplitude error $\delta \Omega = \epsilon \Omega(t)$ or tunable qubit-qubit coupling $J(V)$ susceptible to fluctuations $\delta V$ in the voltage or flux tuning parameter as $\delta J = \delta V \partial_V J(V)$.

By treating $H_\epsilon$ as the interaction Hamiltonian, the solution of the time-dependent Schr\"odinger equation can be expressed as $U(t) = U_\epsilon^\dagger(t) U_c(t)$ where $i \hbar \dot U_c(t) = H_c(t, \boldsymbol p) U_c(t)$ is the ideal time-evolution operator and $i \hbar \dot  U_\epsilon(t) = [U_c^\dagger(t) H_\epsilon(t, \boldsymbol p) U_c(t)] U_\epsilon(t)$ accounts for the effects of the noise. For weak noise ($\int_0^T dt \left|\left| U_c^\dagger(t) H_\epsilon(t, \boldsymbol p) U_c(t)/\hbar\right|\right| \ll \pi$), $U_\epsilon$ can be calculated perturbatively using Magnus expansion as
\begin{align}
U_\epsilon(T) &= \exp\left(-\frac{i}{\hbar}\int_0^T dt U_c^\dagger(t) H_\epsilon(t, \boldsymbol p) U_c(t)\right) + \mathcal O(\epsilon_i^2) \nonumber
\\
&= \exp\left(-\frac{i}{\hbar} \sum_i \epsilon_i \mathcal E_i(T) \right) + \mathcal O(\epsilon_i^2),
\end{align}
where $\mathcal E_i(T) = \int_0^T dt U_c^\dagger(t) \left[\chi_i(t, \boldsymbol p) \Lambda_i\right] U_c(t)$.
A robust quantum gate is one that is insensitive to the first order effects of $\epsilon_i$, i.e. $\partial_{\epsilon_i} U(T)|_{\epsilon_i=0}=0$. This condition can be achieved by choosing a control field for which $\left|\left|\mathcal E_i(T)\right|\right|$ is negligiblly small. The control field must also be chosen such that within the logical subspace $U_c(T)$ has the same result as a given target operation, $U_0$.
These requirements are equivalent to stating that the following cost function needs to vanish:
\begin{multline}
\mathcal C = \left(1- \frac{ \left|\text{tr}\left([\mathcal P U_c(T) \mathcal P^\dagger] U_0^\dagger \right)\right|^k}{D^k}\right)
\\
+ \sum_i \left(\frac{w_i \epsilon_i^\text{max}}{\hbar D}  \mathcal N_{\mathcal P}(\mathcal E_i(T)) \right)^l,
\label{eq:cost}
\end{multline}
where $D \leq n$ is the dimension of the logical subspace used for quantum computation, $\mathcal N_{\mathcal P}(\mathcal E_i(T)) = ||\mathcal P \mathcal E(T) - \text{tr}[\mathcal P \mathcal E(T)]/D||$ is the noise sensitivity of the logical subspace, $\mathcal P$ denotes the projection operator onto the logical subspace,  $||A||^2 =  \text{tr}(A A^\dagger)$, $\epsilon_i^\text{max}$ is the maximum tolerable value of the stochastic noise strength $\epsilon_i$, $w_i \lesssim 1$ are hyperparameters for the optimization step which can be tuned or annealed with a schedule to avoid local minima. The first term in parentheses is similar to the gate infidelity in the absence of stochastic noise which ensures that the final unitary is $U_0$, and the second term is the sensitivity which ensures that the implemented gate is immune to the first order effects of the stochastic noise $\epsilon_i$; this is different from the existing approaches where the cost function is taken to be the noisy gate infidelity which also necessitates an averaging over a particular noise distribution \cite{Nobauer2015,Dong2016,Zahedinejad2016,Bukov2018,Niu2019,Ge2020,Wu2019}. The exponents $k$ and $l$ determine the relative weighting of the fidelity and the sensitivity terms in the cost function. In fact, the two terms could be wrapped inside any monotonically increasing functions, but in our results for simplicity we will simply raise to a power of either 1 or 2.

\begin{figure}
\includegraphics[width=\columnwidth]{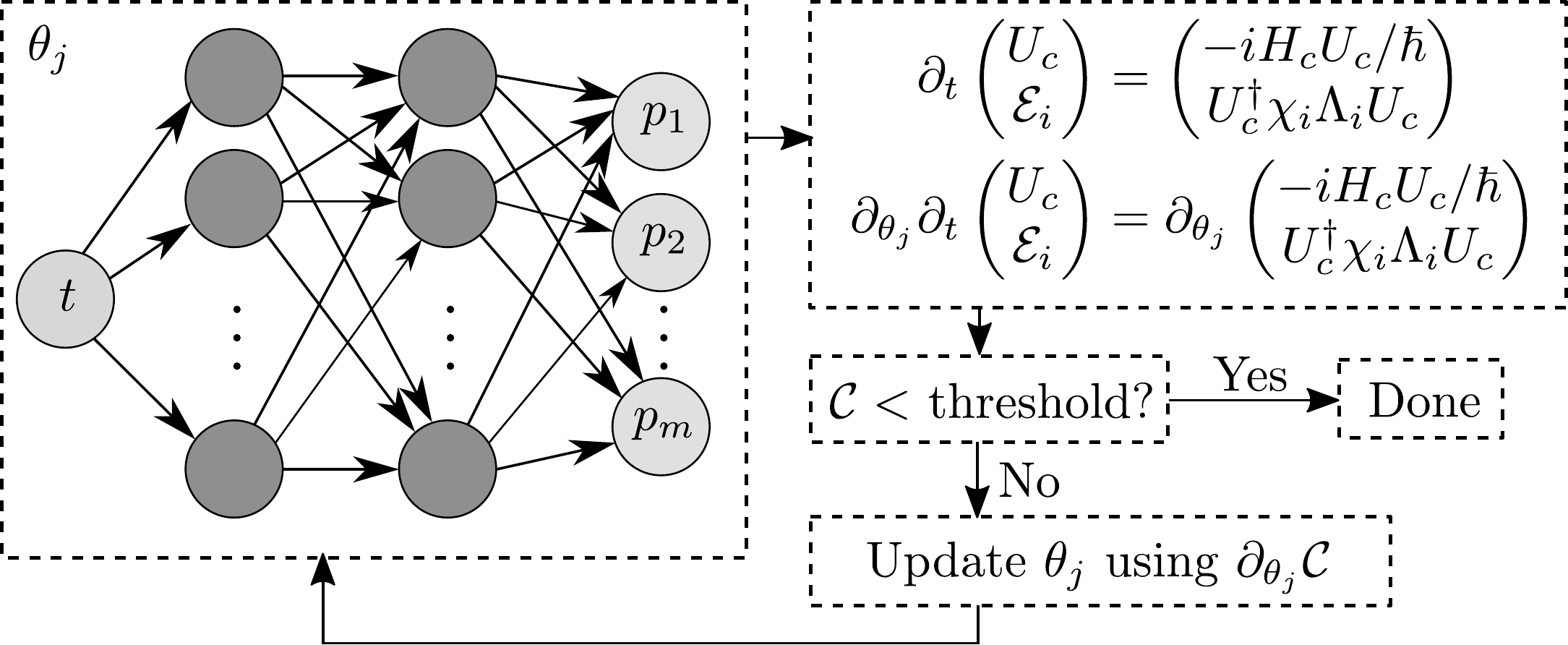}
\caption{(Color online) Optimization flowchart of the physics-informed neural network for finding robust smooth pulses. Backpropagation efficiently works in conjunction with the adjoint sensitivity analysis of the coupled ODE system, and all derivatives with respective to optimization parameters $\theta_j$ (weights, biases and time-scaling parameter) are computed using automatic differentiation \cite{Innes2019}.}
\label{fig:DNN}
\end{figure}

A neural network is a network of ``neurons" (see Fig.~\ref{fig:DNN}), where a vertical column of neurons form a layer. The overall action of the $i$th layer of the neural network is to map a $d_i$ dimensional input vector $\boldsymbol x$ onto a $d_{i+1}$ dimensional output vector $\mathcal L_i(\boldsymbol x) = \sigma_i (\mathcal W_i \boldsymbol x + \boldsymbol b_i)$, where $\mathcal W_i$ is a matrix and $\boldsymbol b_i$ a vector respectively containing the ``weights" and ``biases" of the $i$th layer, and $\sigma_i$ is known as the ``activation function." Our DNN represents a function $t \to \boldsymbol p(t)$, where $\boldsymbol p(t)$ is a smooth curve parameterized by the internal degrees of freedom of the DNN, $\boldsymbol b_i$ and $\mathcal W_i$, as $\boldsymbol p(t) = \mathcal L_N \circ \ldots \circ \mathcal L_1 (t)$.  Intermediate layers $1<i<N$ are commonly referred to as hidden layers. For concreteness, we will take $\sigma_i$ to be the element-wise-acting $\tanh$ function, although functions which are sufficiently  smooth and do not significantly affect the bandwidth requirements of the resulting pulse shape are also viable.

The goal of a DNN optimizer is to vary the weights and biases until the cost function $\mathcal C$ is minimized.
This is achieved by using local gradient based optimization in conjunction with the backpropagation algorithm \cite{Rojas1996}. This requires calculation of $\mathcal C$ in addition to its parameter gradient $\partial \mathcal C$, i.e., the partial derivatives of $\mathcal C$ with respect to each of the elements of $\mathcal W_i$ and $\boldsymbol b_i$ at each iteration, which can be computationally prohibitive.
For this reason, we differentiate $\mathcal E_i(t)$ analytically and transform the problem of calculating $\mathcal C$ into solving a coupled system of ODEs
\begin{align}
\partial_t \begin{pmatrix}
U_c(t) \\
{\mathcal E}_i(t) \\
\end{pmatrix} = \begin{pmatrix}
-i H_c(t, \boldsymbol p) U_c(t)/\hbar \\
U_c^\dagger(t) \left[\chi_i(t, \boldsymbol p) \Lambda_i\right] U_c(t)
\end{pmatrix}
\label{eq:ODE}
\end{align}
subject to the initial conditions $U_c(0) = \openone$, $\mathcal E_i(0) = 0$.
Similarly, $\partial \mathcal C$ is calculated by taking partial derivatives of Eq.~\eqref{eq:ODE} with respect to the elements of $\mathcal W_i$ and $\boldsymbol b_i$ as in Ref.~\cite{Machnes2018}.
This form allows a more efficient calculation of the cost function and its parameter gradient, as required by the backpropagation algorithm, by using the adjoint sensitivity method \cite{Chen2018,Rackauckas2019} on Eq.~\eqref{eq:ODE}. In practice, we perform the straightforward but unwieldy task of calculating the parameter gradients of the coupled ODE system Eq.~\eqref{eq:ODE} by using reverse mode automatic differentiation \cite{Innes2019,Leung2017,Abdelhafez2019}.

The above analysis can be extended to higher order error correction, which is relevant when the noise is not sufficiently weak for a first order treatment. For example, second order error correction can be achieved by appending $\partial_t {\mathcal E_{ij}^{(2)}}(t) =  [\partial_t \mathcal E_i(t) \mathcal E_j(t) - \mathcal E_j(t) \partial_t \mathcal E_i(t)]/2$ to Eq.~\eqref{eq:ODE} with the initial condition ${\mathcal E_{ij}^{(2)}}(0)=0$, and introducing $\sum_{i,j}  (w_{ij} \epsilon_i^\text{max} \epsilon_j^\text{max} \mathcal N_{\mathcal P} (\mathcal E_{ij}^{(2)}(T)) /D \hbar^2)^l $ to the cost function, which ensures that the second order term in the Magnus expansion for $U_\epsilon(T)$ vanishes, and comes at only minor additional computational cost.
\footnote{Similarly, third order error correction can be enforced using $\partial_t \mathcal E_{ijk}^{(3)}(t) = [\mathcal E_{jk}^{(2)}(t) \partial_t\mathcal E_i(t) -\mathcal E_j(t)\partial_t\mathcal E_i(t) \mathcal E_k(t) - \mathcal E_k(t)\partial_t\mathcal E_i(t) \mathcal E_j(t) + \partial_t\mathcal E_i(t) \mathcal E_{kj}^{(2)}(t)]/3$, and so on.}

We remark that we are performing a search in the space of functions. In general, a large number of parameters are required to adequately explore a functional space. Parametrizing a functional space in a way that leads to convergent and experimentally feasible solutions is a nontrivial problem \cite{Barnes2015,Zeng2019,Gungordu2019b,Gungordu2020a}. For example, a set of a few sinusoidal harmonics \cite{Caneva2011,Nobauer2015,Machnes2018} explores a very limited portion of the functional space; this can be remedied by increasing the number of harmonics but results in unrealistic bandwidth requirements and significantly slows down the optimization problem due to inefficient computation of gradients and reduced timesteps in adaptive ODE solvers.

In contrast, our parametrization of the time-dependence of the Hamiltonian eliminates the problem of finding a well-behaved hand-crafted ansatz form to optimize over, and makes it straightforward to explore a larger functional space by simply increasing the number of neurons and layers, owing to the fact that DNNs are universal function approximators \cite{universality}, and produces well-behaved Hamiltonians while remaining computationally efficient.
We emphasize that our parameterization with DNNs, $t \to \boldsymbol p(t) \to H_c(t, \boldsymbol p)$, differs from earlier works: instead of using the output of the DNN to represent discrete ansatz parameters (such as the amplitudes of harmonics \cite{Nobauer2015} or segments of piecewise control fields \cite{Dong2016,Zahedinejad2016,Palittapongarnpim2017,Bukov2018,Niu2019,Ge2020,Wu2019}), we directly use $\boldsymbol p(t)$ as a smooth differentiable function of time to construct $H_c(t)$, which is made possible by the recently introduced PINNs \cite{Raissi2019}. Our approach has the added benefit that functional constraints can be specified as a part of the parametrization $\boldsymbol p(t) \to H_c(t,\boldsymbol p)$ (e.g., $H_c(t, \boldsymbol p) = A \sin(p_1(t))\sigma_x + B \sigma_z$ constrains the drive amplitude to be less than $|A|$) which limits the search to the space of functions that satisfy the constraints from the outset. This is in contrast to enforcing constraints via cost function by penalizing any deviations, which complicates the optimization landscape and can introduce false minima, leading to inefficiency.

Since both the adjoint sensitivity method and backpropagation algorithm (which uses the directed graph structure of DNNs to compute the gradients more efficiently) are well established computational methods, we here focus on presenting our results on robust quantum control, and refer the reader to the literature \cite{Rojas1996,Chen2018,Rackauckas2019} for their details.
For our numerical results, we use DiffEqFlux.jl \cite{Rackauckas2019}, which is a Julia package implementing a PINN optimizer, with optional support for hardware DNN accelerators.
Before moving on to explicit solutions, we remark that the gate time $T$ needs to be decided prior to optimization. For Hamiltonians that can contain non-adjustable terms (such as drift terms), a solution will generally not exist for an arbitrary $T$, and finding a suitable value can be laborious. This problem can be solved by introducing a new optimization parameter $\alpha$, a time-independent scaling factor, to the right hand side of Eq.~\eqref{eq:ODE} and to the noise terms $\epsilon_i$ in Eq.~\eqref{eq:cost};
  $\alpha$ can be seen as a scaling factor for time or the total Hamiltonian. To avoid long gate times approaching to inverse of the incoherent decay rates $\Gamma_i$, one can also introduce a factor $e^{-\sum_i \Gamma_i T / \alpha}$ to the trace fidelity term in the cost function $\mathcal C$.

\begin{figure}
\includegraphics[width=0.9\columnwidth]{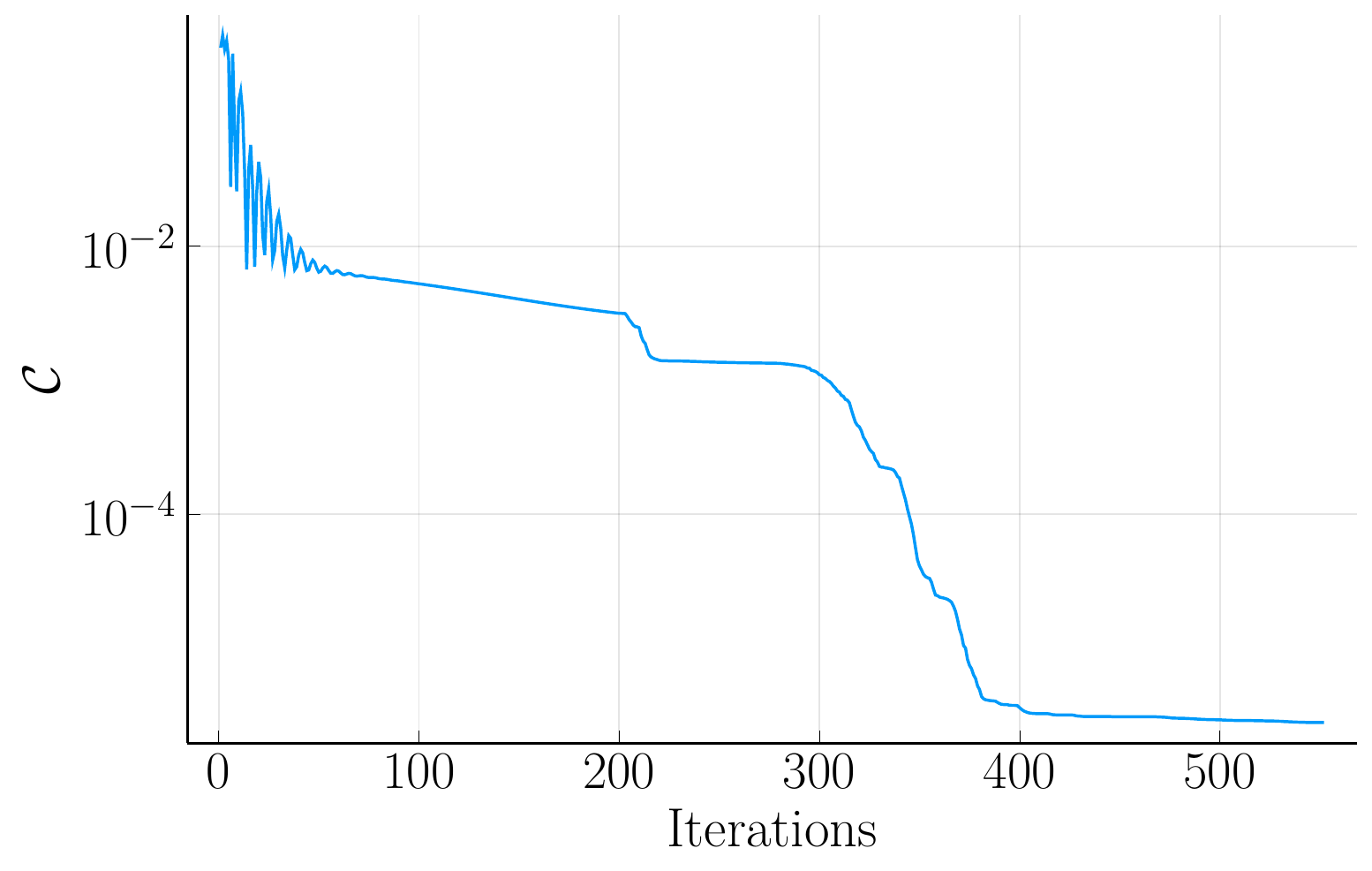}
\caption{(Color online) Decay of the cost function $\mathcal C$ as a function of iterations during the optimization process for the pulse shown in Fig.~\ref{fig:transmon}.}
\label{fig:transmon-cost}
\end{figure}
To train the neural networks, we use local gradient based optimization algorithms. We start from a random initial internal state for the neural network, and use a variant of stochastic gradient descent algorithm with limited iterations, followed by Broyden-Fletcher-Goldfarb-Shanno (BFGS) passes.
A representative plot for the decay of the cost function is shown in Fig.~\ref{fig:transmon-cost}, which was obtained during the training process for the pulse shown in Fig.~\ref{fig:transmon}. During the first 200 iterations, AMSGrad \cite*{amsgrad} was used with the learning rate $\eta=10^{-3}$. The result was further refined using a BFGS pass of 350 iterations with an initial step norm of $10^{-3}$.
The fully connected DNN layer structure with two hidden layers of length 32 provides $\sim 10^3$ optimization parameters, which we found to be sufficient for all the problems we have studied in this work.

\section{Examples}

\subsection{Exchange-coupled spin qubits}
\begin{figure}
\includegraphics[width=0.75\columnwidth]{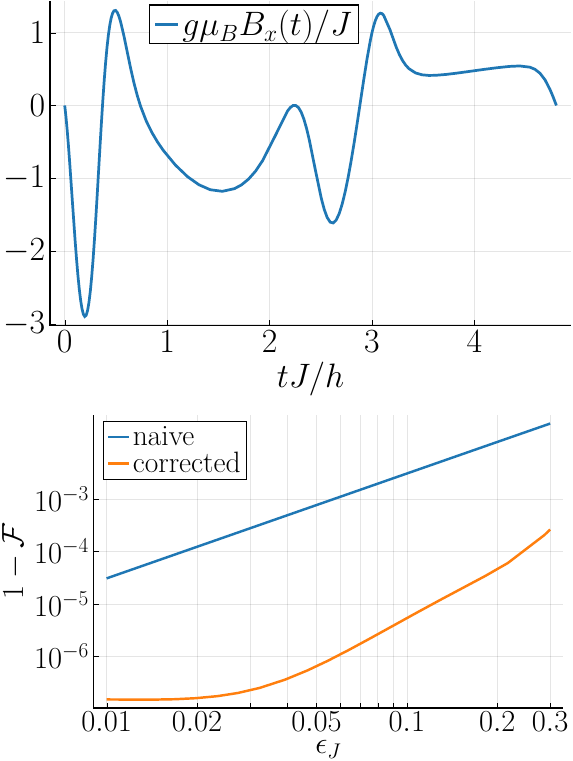}
\caption{(Color online) (Top) Pulse shape implementing a CZ gate that is robust against the first and second order effects of quasistatic errors in exchange coupling, obtained with $\kappa T=30$,  $k=2$, $l=2$, $T=4.8 h/J$ using a neural network with two hidden layers of length 32, in $\approx 1400$ optimization steps. The detailed numerical parameters of this pulse are tabulated in Appendix \ref{sec:parameters}. (Bottom) Gate infidelity as a function of the error strength $\epsilon_J$ for the robust pulse shape (corrected) and a simple, undriven implementation (naive).}
\label{fig:dqd}
\end{figure}

As our first proof-of-concept example, we consider a pair of spin qubits in a semiconductor double quantum dot, with one electron in each dot. The overlap between the electron wavefunctions is determined by the gate voltages, which provides voltage tunable exchange coupling between the spin degrees of freedom of the electrons. Single-qubit operations are realized by modulation of the magnetic field that is generated by an on-chip wire. In a rotating frame, the spin Hamiltonian of this system can be written as \cite{Gungordu2020a}
\begin{align}
H_c(t) = \frac{J}{4}\sigma_z \otimes \sigma_z + \frac{1}{2} g \mu_B B_x (t) \sigma_x \otimes \openone,
\end{align}
whose Lie algebra is isomorphic to a two-level $\mathfrak{su}(2)$ generated by $\{\sigma_z \otimes \sigma_z, \sigma_x \otimes \openone, \sigma_y \otimes \sigma_z\}$. The exchange coupling, $J$, which in some devices is essentially fixed due to bandwidth limitations \cite{Huang2019,Gungordu2020a}, is susceptible to charge noise induced fluctuations, which can be modelled by the noise Hamiltonian $H_\epsilon = \epsilon_J J \sigma_z \otimes \sigma_z/4$.
An entangling CZ-equivalent gate ($e^{-i \sigma_z \otimes \sigma_z \pi/4}$) can be produced naively by setting $B_x = 0$ and waiting a time $\hbar \pi/J$.  However, instead we search for a smooth pulse $B_x(t)$ which can correct both first and second order quasistatic fluctuations in $J$ by parameterizing the magnetic field as $g \mu_B B_x(t)/2 = (J/4) A(t) p_1(t) \sin(p_2(t))$. Here, $A(t) = \coth(\kappa T) [\tanh(\kappa t) - \tanh(\kappa (t-T))] - 1$ is a smoothed unit square pulse ($\kappa$ determines the degree of the smoothing), and its purpose is to enforce the condition that the magnetic field is turned on only for the duration of the pulse, $B_x(0) = B_x(T) = 0$. The resulting smooth pulse shown in Fig.~\ref{fig:dqd} produces a CZ gate with extraordinarily high gate fidelities (defined as $\mathcal F = |\text{tr}(U(T) U_0^\dagger)/4|^2$) above 99.99\% for exchange errors as large as 24\%, and is faster than the smooth pulse reported in Ref. \cite{Barnes2012}. A bandwidth limitation of $\Delta f \approx 20/T$ leads an infidelity $\approx 10^{-5}$, which corresponds to a very conservative value of $\approx 4$MHz for $J/h=1$MHz \footnote{Incidentally, quasistatic $\sigma_z \otimes \openone$ errors, which could be due to residual spinful isotopes in the substrate or miscalibration, are also corrected due to the structure of the algebra embedding \cite{Gungordu2020a}.}.

\begin{figure}
\includegraphics[width=0.75\columnwidth]{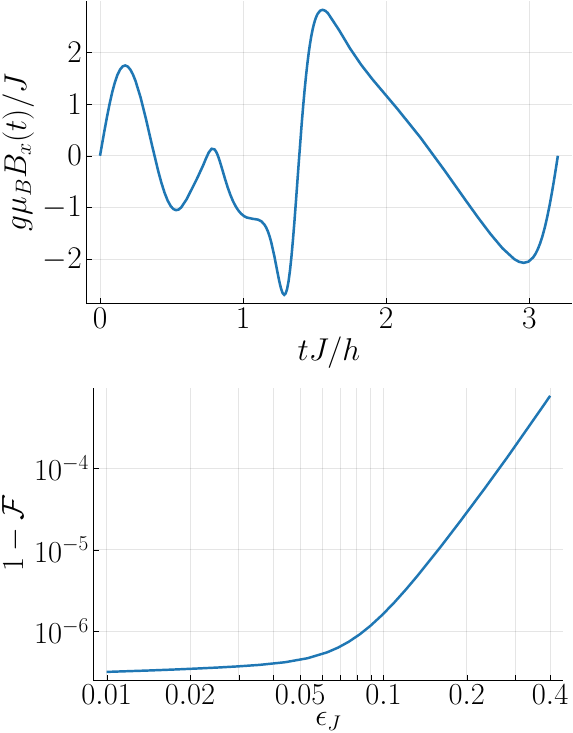}
\caption{(Color online) (Top) Pulse shape implementing a $X_{\pi/2}$ gate that is robust against the first order effects of quasistatic errors in exchange coupling, obtained with $\kappa T=20$,  $k=2$, $l=2$, $T=3.2 h/J$ using a neural network with two hidden layers of length 32. The detailed numerical parameters of this pulse are given in the text. (Bottom) Gate infidelity as a function of the error strength $\epsilon_J$. No infidelity curve for a naive implementation is shown, because one-qubit gates with always-on $J$
coupling are a nontrivial problem even without any robustness requirements \cite*{Huang2019}.}
\label{fig:dqd-x}
\end{figure}

For completeness, we now turn to full SU(4) control in such systems. A robust universal set of gates in exchange-coupled spin qubits can be achieved with the addition of error-free virtual Z rotations \cite{McKay2017} and the robust one-qubit rotation $X_{\pi/2}$ \cite{Gungordu2020a} implemented by the pulse shown in Fig.~\ref{fig:dqd-x}. Either qubits can be targeted by changing the modulation frequency of the driving field to that of the target qubit \cite{Gungordu2020a}.

\begin{figure}
\includegraphics[width=0.75\columnwidth]{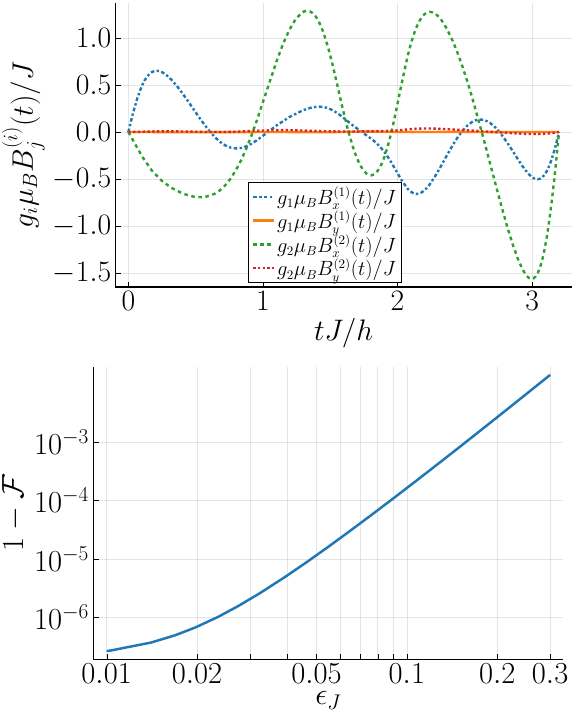}
\caption{(Color online) (Top) Pulse shape implementing the gate $\exp\left(-i\frac{\pi}{8}[\sigma_y \otimes \sigma_y + \sigma_z \otimes \sigma_z]\right)$ that is robust against the first order effects of quasistatic errors in exchange coupling, obtained with $\kappa T=20$,  $k=2$, $l=2$, $T=3.2 h/J$ using a neural network with two hidden layers of length 32. The detailed numerical parameters of this pulse are given in the text. (Bottom) Gate infidelity as a function of the error strength $\epsilon_J$. No infidelity curve for a naive implementation is shown, because implementation of an $\sqrt{\text{iSWAP}}$ gate with always-on $J$ coupling is a nontrivial problem even without any robustness requirements as it requires one-qubit rotations with always-on $J$.}
\label{fig:dqd-iswap}
\end{figure}

We can also use this method to find robust pulses for arbitrary elements of $\text{SU}(2)\subset \text{SU}(4)$ generated by $\{\sigma_z \otimes \sigma_z, \sigma_x \otimes \openone, \sigma_y \otimes \sigma_z \}$ or  $\{\sigma_z \otimes \sigma_z,  \openone \otimes \sigma_x, \sigma_z \otimes \sigma_y \}$, in a direct manner provided that the gate time is sufficiently long (for example, an entangling operation requires at least $T \gtrsim \hbar \pi/4J$). Arbitrary SU(4) rotations can also be implemented using a two-tone drive, given by
\begin{align}
H_c(t) = &\frac{J}{4}\sigma_z \otimes \sigma_z + \frac{1}{2} g_1 \mu_B [ B_x^{(1)} (t) \sigma_x \otimes \openone + B_y^{(1)} (t) \sigma_y \otimes \openone] + \nonumber \\
&\frac{1}{2} g_2 \mu_B [ B_x^{(2)} (t) \openone \otimes \sigma_x + B_y^{(2)} (t) \openone \otimes \sigma_y],
\end{align}
The generators of this Hamiltonian fully span $\mathfrak{su}(4)$, and hence can generate any SU(4) gate directly. As a representative example, the pulse shown in Fig.~\ref{fig:dqd-iswap} generates $\exp\left(-i\frac{\pi}{8}[\sigma_y \otimes \sigma_y + \sigma_z \otimes \sigma_z]\right)$ which is equivalent to an $\sqrt{\text{iSWAP}}$ gate up to local unitaries. We note that this rotation cannot be generated directly using the Hamiltonian in Eq. (5).

\subsection{Transmon qubit}
To illustrate simultaneous noise and leakage suppression, we next turn to the transmon qubit, although we remark in passing that this scenario is also relevant in the context of resonator-coupled spin qubits \cite{Warren2019} and encoded exchange-only spin qubits \cite{Russ2017}.
The effective Hamiltonian for the transmon can be written as \cite{Gambetta2011}
\begin{align}
H_c(t) \approx \delta(t) a^\dagger a + \frac{\Delta}{2} a^\dagger a(a^\dagger a-\openone) + \frac{\Omega(t) a + \Omega^*(t) a^\dagger}{2}
\end{align}
in a rotating frame, where $\Delta$ is the anharmonicity and $\Omega(t)$ is the complex envelope of the microwave drive capacitively coupled to the transmon via a resonator, whose frequency is detuned from the qubit frequency by $\delta(t)$. The first two levels encode a logical qubit, and we consider the first four levels in our calculations \footnote{All results were later reaffirmed using 10 levels.}.
Thus, when calculating $\mathcal C$, we project $U(T)$ onto the qubit subspace. The single-sided projection $\mathcal P \mathcal E_i(T)$ in $\mathcal N_{\mathcal P}(\mathcal E_i(T))$ allows us to leave out the effects of the noise solely on leakage subspace \footnote{If the input states are guaranteed to be fully within in the logical subspace, the sensitivity cost can be simplified to $||\mathcal P \mathcal E_i(T)\mathcal P^\dagger||$}; that being said, it is possible to protect the leakage subspace as well by simply omitting this projection. Our goal is to find a smooth pulse for implementing a gate that can suppress both leakage and shifts in detuning $\delta(t) \to \delta(t) + \epsilon$ which can be caused by calibration errors \cite{Lucero2010}, stochastic phase errors \cite{Ball2016}.

At this point, we recall that an established method of suppressing leakage in Josephson junction based qubits is to use pulse shapes that obey a particular family of differential relations between $\Omega(t)$, $\Delta$ and $\delta(t)$, known as DRAG \cite{Motzoi2009}. These relations ensure that the leakage inducing terms remain small throughout the pulse. It is possible to enforce DRAG conditions by construction, for instance by augmenting Eq.~\eqref{eq:ODE} with the relation $\dot\Omega_P(t) = p_1(t) \sin(p_2(t))$ and parameterizing the drive as $\Omega(t) = A(t) \Omega_P(t) - i \partial_t [A(t) \Omega_P(t)]/2\Delta$, $\delta(t)=0$. However, we will proceed without doing so in order to avoid limiting the search space: DRAG is a sufficient condition for suppressing leakage, but it is not a necessary one since what matters is whether the qubit subspace time-evolution operator is equal to the target unitary at the final time $t=T$, regardless of any leakage that may be present during intermediate times $0 < t < T$.
\begin{figure}
\includegraphics[width=0.75\columnwidth]{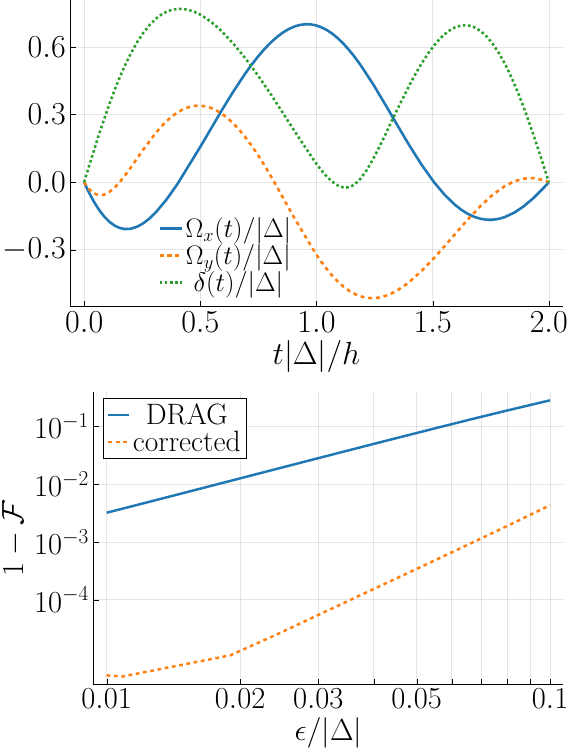}
\caption{(Color online) (Top) Pulse shape implementing a $X_{\pi/2}$ gate that is robust against the first order effects of quasistatic errors in detuning and leakage errors, obtained with $\kappa T=50/4$, $k=1$, $l=2$, $T=2h/|\Delta|$ using a neural network with two hidden layers of length 32, in $\approx 600$ optimization iterations. The detailed numerical parameters of this pulse are tabulated in Appendix \ref{sec:parameters}. (Bottom) Gate infidelity as a function of the error strength $\epsilon/|\Delta|$.}
\label{fig:transmon}
\end{figure}
We thus parameterize the driving field as
\begin{align}
\Omega(t) =& \Omega_x(t) + i\Omega_y(t)  \nonumber\\
=& 4\Delta A(t)(2/\pi)[\arctan(p_1(t)) \sin(p_2(t)) + \nonumber\\
& i \arctan(p_3(t)) \sin(p_4(t))], \nonumber\\
\delta(t) =& 2\Delta A(t)(2/\pi)\arctan(p_5(t)) \sin(p_6(t)),
\end{align}
where the use of $\arctan$ clamps the field amplitudes, and we target a $X_{\pi/2}$ gate \footnote{When it is a concern, one can incorporate filtering effects (modeled e.g. by a Hann filter) into the optimization procedure by changing the parameterization as  $\delta(t) = \mathfrak F [2\Delta A(t)(2/\pi)\arctan(p_5(t)) \sin(p_6(t)]$ and similarly for $\Omega(t)$.}. This is in particular a useful example, because when used together with error-free virtual $Z$ rotations implemented by shifting the rotating frame \cite{McKay2017},
$X_{\pi/2}$ gates are sufficient to implement arbitrary single qubit rotations. The resulting pulse shape is shown in Fig.~\ref{fig:transmon}.
For a typical anharmonicity value of $\Delta/h \sim -200$MHz \cite{Gambetta2011},
the pulse duration is $T \sim 10$ns.
Under this assumption, the resulting gate fidelities, defined by $\mathcal F = |\text{tr}([\mathcal P U(T) \mathcal P^\dagger] U_0^\dagger)/2|^2$, remain above 99.99\% as long as the shift in detuning remains below 3.5\% of $\Delta$.  When the effect of two additional higher leakage states are taken into account for this pulse shape, the baseline fidelity remains the same. Compared to a nonrobust pulse based on DRAG \cite{Motzoi2009,Lucero2010,Gambetta2011}, which can take at least $T\approx 2.1h/|\Delta|=10.5$ns and reaches the same infidelity threshold at 0.03\% of $\Delta$, this new shaped pulse improves the error threshold against detuning errors by two orders of magnitude. A bandwidth limitation $\Delta f \approx 4.73/T$ leads to $10^{-5}$ infidelity, which corresponds to $\approx 473$MHz for $\Delta=-200$MHz, approximately twice the bandwidth required by DRAG $\approx 255$MHz.

Although dephasing effects associated with quasistatic fluctuations are corrected by a robust pulse, relaxation processes can limit the fidelity of gate operations in transmon. These effects can be quantified using the master equation
\begin{align}
\dot \rho(t) = -\frac{i}{\hbar} [H(t), \rho(t)]+\frac{1}{T_1} \mathcal D[a]\rho(t) + \frac{1}{T_\phi} \mathcal D[a^\dagger a]\rho(t)
\end{align}
where $\mathcal D[A]\rho(t) \equiv A \rho(t) A^\dagger - \{A^\dagger A,\rho(t)\}/2$ and $1/T_\phi = 1/T_2 - 1/2T_1$ is the pure dephasing rate. We compute the state-averaged gate fidelity \cite{Cabrera2007a}
\begin{align}
\langle\mathcal F\rangle = \frac{1}{2} + \frac{1}{(2+1)2^2}\sum_{i=1}^3 \text{tr}[\mathcal Q (\sigma_i) U(T)\sigma_i U(T)^\dagger]
\end{align}
where $\mathcal Q$ represents the quantum channel associated with the master equation as $\rho_0 \to \mathcal Q(\rho_0)$.
For this robust pulse, we find that the effects of relaxation on the fidelity is less than $10^{-4}$ when $T_1 |\Delta| / h \geq 3554$, and $T_\phi \geq 10 T_1$ (we remark, however, that this represents a higher bar than necessary, given that our pulse shape already protects against pure quasistatic dephasing errors). These limits are well accessible in recent experiments; for example in Ref. \cite{Sung2021}, the value of $T_1 |\Delta|/h$ is 13200 with $|\Delta|/h = 220$MHz, $T_1=60\mu$s and $T_\phi \gtrsim 727\mu\text{s}$ (lower limit, obtained from $T_2^\text{Hahn}=103\mu$s which is smaller than $T_2$) which is greater than $10T_1$.

 We remark that when drive amplitude limitations or the robustness constraints are removed, our method easily produces faster nonrobust gates that well outperform the fastest DRAG pulses, ultimately limited by the bandwidth of the drive.
More stringent experimental bandwidth constraints can be accommodated by running the search with an appropriately increased gate time $T$; although it is harder to find solutions for shorter gate times, one can always find solutions at longer times.

\section{Conclusion}
In summary, we have introduced a method for performing dynamically corrected quantum gates with practical smooth pulses that is broadly applicable to large Hilbert spaces beyond two-level systems, which is necessary for robust control of multi-qubit devices and  leakage into excited states, and is practically extensible to correction of errors beyond first order. Our approach is the first generally applicable robust noise-sensitivity based smooth pulse shaping method, does not require sampling or assumption with regards to the nature of the noise, and leverages physics-informed deep neural networks for computational advantages. In addition to noise cancellation, our generic approach can also be used to reduce or eliminate the need for careful recalibration cycles during experiments. A future direction is the extension of this approach to suppress time-dependent broadband noise.
\begin{acknowledgments}
This research was sponsored by the Army Research Office (ARO), and was accomplished under Grant Number W911NF-17-1-0287.
\end{acknowledgments}

\appendix

\section{Parameters for shaped pulses in the main text}
\label{sec:parameters}
The dimensionless pulse shapes shown in the main text can be approximated using a truncated Chebyshev series in the form
\begin{align}
\sum_{n=0}^{N-1} c_n T_n(2\tau/\tau_0-1),
\end{align}
where $T_n(x)$ is $n$th the Chebyshev polynomial of the first kind, $\tau$ ($\tau_0$) is the dimensionless (total) time shown along the $x$-axis. The Chebyshev coefficients $c_n$ for $g \mu_B B_x(\tau)/J$ shown in Fig.~\ref{fig:dqd} are tabulated in Table \ref{tab:dqd}
Similarly, the Chebyshev series coefficients for $g \mu_B  B_x(\tau)/J$ shown in Fig.~\ref{fig:dqd-x} are tabulated in Table \ref{tab:dqd-x}, and for $g_1 \mu_B  B_x^{(1)} (\tau)/J$, $g_1 \mu_B  B_y^{(1)} (\tau)/J$, $g_2 \mu_B  B_x^{(2)} (\tau)/J$ and $g_2 \mu_B  B_y^{(2)} (\tau)/J$ in Fig.~\ref{fig:dqd-iswap} are tabulated in Table \ref{tab:dqd-iswap}. Finally, for $\Omega_x(\tau)/\Delta$ $\Omega_y(\tau)/\Delta$, and $\delta(\tau)/\Delta$ in Table \ref{tab:transmon}. Approximate pulse shapes reconstructed from these Chebyshev coefficients result in similar fidelity curves which reach the $10^{-4}$ threshold around the same value, closely approximating the fidelity curves given in the main text. The code used to produce the robust pulses shown in Fig.~\ref{fig:dqd} and Fig.~\ref{fig:transmon}, and the resulting internal state of the neural networks can be found in the supplemental files \onlinecite{supplement}.

\begin{table}
\begin{tabular}{|c|c|}\toprule
$n$ & $g \mu_B B_x(\tau)/J$ \\
\hline
0 & -0.222\\
1 & 0.714\\
2 & 0.0125\\
3 & -0.178\\
4 & -0.3635\\
5 & 0.178\\
6 & 0.1255\\
7 & -0.3785\\
8 & 0.3445\\
9 & -0.556\\
10 & 0.433\\
11 & -0.012\\
12 & -0.0975\\
13 & -0.047\\
14 & -0.164\\
15 & 0.406\\
16 & -0.145\\
17 & -0.107\\
18 & -0.0315\\
19 & 0.076\\
20 & 0.1425\\
21 & -0.1545\\
22 & -0.034\\
23 & 0.044\\
24 & 0.0415\\
25 & 0.008\\
26 & -0.065\\
27 & 0.017\\
28 & 0.016\\
\hline
\end{tabular}
\caption{Chebyshev series expansion coefficients $c_n$ for $g \mu_B B_x(\tau)/J$ with $\tau_0 = 4.8$ for the pulse shown in Fig.~\ref{fig:dqd}.}
\label{tab:dqd}
\end{table}

\begin{table}
\begin{tabular}{|c|c|}\toprule
$n$ & $g \mu_B B_x(\tau)/J$ \\
\hline
0 & -0.221 \\
1 & -0.339 \\
2 & -0.617 \\
3 & -0.422 \\
4 & 0.576 \\
5 & 0.571 \\
6 & 0.039 \\
7 & 0.248 \\
8 & 0.347 \\
9 & -0.131 \\
10 & -0.119 \\
11 & 0.189 \\
12 & -0.019 \\
13 & -0.167 \\
14 & 0.066 \\
15 & 0.072 \\
16 & -0.1 \\
17 & -0.015 \\
18 & 0.076 \\
19 & -0.027 \\
20 & -0.042 \\
21 & 0.04 \\
22 & 0.013 \\
23 & -0.033 \\
24 & 0.007 \\
25 & 0.021 \\
26 & -0.014 \\
27 & -0.008 \\
28 & 0.013 \\
29 & -0.001 \\
30 & -0.009 \\
31 & 0.004 \\
32 & 0.004 \\
33 & -0.005 \\
34 & -0.001 \\
35 & 0.004 \\
36 & -0.001 \\
37 & -0.002 \\
38 & 0.002\\
\hline
\end{tabular}
\caption{Chebyshev series expansion coefficients $c_n$ for $g \mu_B B_x(\tau)/J$ with $\tau_0 = 3.2$ for the pulse shown in Fig. \ref{fig:dqd-x}.}
\label{tab:dqd-x}
\end{table}

\begin{table}
\begin{tabular}{|c|c|c|c|c|}\toprule
$n$ & $\frac{ g_1 \mu_B  B_x^{(1)} (\tau)}{J}$ & $\frac{g_1 \mu_B  B_y^{(1)} (\tau)}{J}$ & $\frac{g_2 \mu_B  B_x^{(2)} (\tau)}{J}$ & $\frac{g_2 \mu_B  B_y^{(2)} (\tau)}{J}$\\
\hline
0 & 0.001 & 0 & -0.156 & 0.005\\
1 & -0.312 & 0 & -0.151 & -0.004\\
2 & 0.066 & 0 & -0.594 & -0.012\\
3 & 0.036 & -0.001 & -0.241 & -0.005\\
4 & -0.035 & 0.001 & 0.406 & 0.001\\
5 & 0.056 & 0 & 0.121 & 0.007\\
6 & -0.124 & 0 & 0.416 & 0.007\\
7 & 0.262 & 0 & 0.445 & 0.006\\
8 & 0.132 & 0 & -0.113 & -0.003\\
9 & -0.016 & 0 & -0.282 & -0.003\\
10 & -0.045 & 0 & 0.09 & 0.003\\
11 & -0.038 & -0.001 & 0.191 & 0.001\\
12 & -0.001 & 0 & -0.045 & -0.001\\
13 & 0.027 & 0.001 & -0.119 & 0\\
14 & 0.015 & 0 & -0.008 & -0.001\\
15 & -0.018 & -0.001 & 0.056 & 0\\
16 & -0.016 & 0 & 0.014 & 0.001\\
17 & -0.002 & 0 & -0.028 & 0\\
18 & 0.012 & 0 & -0.017 & -0.001\\
19 & 0.008 & 0 & 0.008 & 0\\
20 & -0.008 & 0 & 0.011 & 0\\
21 & -0.009 & 0 & 0 & 0\\
22 & 0.003 & 0 & 0 & 0\\
23 & 0.008 & 0 & 0 & 0\\
24 & 0 & 0 & 0 & 0\\
25 & -0.005 & 0 & 0 & 0\\
26 & -0.002 & 0 & 0 & 0\\
27 & 0.003 & 0 & 0 & 0\\
28 & 0.002 & 0 & 0 & 0\\
\hline
\end{tabular}
\caption{Chebyshev series expansion coefficients $c_n$ for $g_i \mu_B B_j(\tau)/J$ with $\tau_0 = 3.2$ for the pulse shown in Fig.~\ref{fig:dqd-iswap}.}
\label{tab:dqd-iswap}
\end{table}

\begin{table}
\begin{tabular}{|c|c|c|c|}\toprule
$n$ & $\Omega_x(\tau)/\Delta$ & $\Omega_y(\tau)/\Delta$ & $\delta(\tau)/\Delta$ \\
\hline
0 & 0.004 & -0.0153 & 0.3267\\
1 & -0.0216 & -0.0728 & -0.0696\\
2 & -0.181 & 0.0997 & -0.2557\\
3 & 0.1319 & 0.2563 & 0.0677\\
4 & 0.2608 & -0.1326 & -0.0823\\
5 & -0.1138 & -0.2096 & 0.0041\\
6 & -0.1005 & 0.0578 & 0.0087\\
7 & -0.0053 & 0.0243 & -0.0022\\
8 & 0.018 & -0.0104 & 0.0028\\
9 & 0.0112 & 0.002 & 0\\
10 & -0.0011 & 0 & 0\\
11 & -0.0027 & 0 & 0\\
\hline
\end{tabular}
\caption{Chebyshev series expansion coefficients $c_n$ for $\Omega_x(\tau)/\Delta$, $\Omega_y(\tau)/\Delta$ and $\delta(\tau)/\Delta$ with $\tau_0 = 2$ for the pulse shown in Fig.~\ref{fig:transmon}.}
\label{tab:transmon}
\end{table}

\bibliography{siqd,extra,dnn,transmon,resonator}

\end{document}